\documentclass[aps,prb,twocolumn,superscriptaddress,showpacs,floatfix]{revtex4-1}
\usepackage{graphicx}
\usepackage{epstopdf}
\usepackage{color}
\definecolor{orange}{RGB}{255,127,0}
\definecolor{blue2}{RGB}{33,114,173}

\begin{document}

% Use the \preprint command to place your local institutional report
% number in the upper righthand corner of the title page in preprint mode.
% Multiple \preprint commands are allowed.
% Use the 'preprintnumbers' class option to override journal defaults
% to display numbers if necessary
%\preprint{}
%Title of paper
\title{Making the Dzyaloshinskii-Moriya interaction visible}

\author{A.~Hrabec}
\affiliation{Laboratoire de Physique des Solides, Univ. Paris-Sud, Universit\'{e} Paris-Saclay, CNRS, UMR 8502, 91405 Orsay Cedex, France}
%\affiliation{Laboratoire Charles Coulomb, Universit\'{e} de Montpellier and CNRS UMR 5221, 34095 Montpellier, France}
\author{M.~Belmeguenai}
\affiliation{LSPM (CNRS-UPR 3407), Universit\'{e} Paris 13, Sorbonne Paris Cit\'{e}, 99 avenue Jean-Baptiste Cl\'{e}ment, 93430 Villetaneuse, France}
\author{A.~Stashkevich}
\affiliation{LSPM (CNRS-UPR 3407), Universit\'{e} Paris 13, Sorbonne Paris Cit\'{e}, 99 avenue Jean-Baptiste Cl\'{e}ment, 93430 Villetaneuse, France}
\author{S.M.~Ch\'{e}rif}
\affiliation{LSPM (CNRS-UPR 3407), Universit\'{e} Paris 13, Sorbonne Paris Cit\'{e}, 99 avenue Jean-Baptiste Cl\'{e}ment, 93430 Villetaneuse, France}
\author{S.~Rohart}
\affiliation{Laboratoire de Physique des Solides, Univ. Paris-Sud, Universit\'{e} Paris-Saclay, CNRS, UMR 8502, 91405 Orsay Cedex, France}
\author{Y.~Roussign\'{e}}
\affiliation{LSPM (CNRS-UPR 3407), Universit\'{e} Paris 13, Sorbonne Paris Cit\'{e}, 99 avenue Jean-Baptiste Cl\'{e}ment, 93430 Villetaneuse, France}
\author{A.~Thiaville}
\affiliation{Laboratoire de Physique des Solides, Univ. Paris-Sud, Universit\'{e} Paris-Saclay, CNRS, UMR 8502, 91405 Orsay Cedex, France}
\email[]{andre.thiaville@u-psud.fr}

\begin{abstract}
Brillouin light spectroscopy is a powerful and robust technique for measuring the interfacial Dzyaloshinskii-Moriya interaction in thin films with broken inversion symmetry. Here we show that the magnon visibility, i.e. the intensity of the inelastically scattered light, strongly depends on the thickness of the dielectric seed material - SiO$_2$. By using both, analytical thin-film optics and numerical calculations, we reproduce the experimental data. We therefore provide a guideline for the maximization of the signal by adapting the substrate properties to the geometry of the measurement. Such a boost-up of the signal eases the magnon visualization in ultrathin magnetic films, speeds-up the measurement and increases the reliability of the data.
\end{abstract}
\pacs{75.70.Tj; 75.76.+j; 75.50.-y}
\date{\today}

\maketitle
Magnons, quanta of spin waves\cite{bloch1930theorie}, carrying a fixed angular momentum have been proposed to be used in the field of spintronics due to their ability of transporting an information in a Joule heat-free manner. They are typically detected electrically\cite{owens1985system}, optically by using X-rays\cite{wintz2016magnetic} or via the neutron\cite{mook1985neutron} or Brillouin light scattering (BLS) mechanisms\cite{demokritov2001brillouin}. The most commonly used media to transmit the magnons over large distances are metallic films of Permalloy\cite{kruglyak2010magnonics} or the low-damping insulator yttrium--iron--garnet\cite{cherepanov1993saga}. However, with the advent of ultrathin magnetic films, where the interfacial effects can be beneficial to magnon spintronics (magnonics), the conventional detection techniques reach their limits due to a small probed volume.

In ultrathin films with broken inversion symmetry the Dzyaloshinskii-Moriya interaction (DMI) can be present, causing a non-reciprocity of spin wave propagation\cite{udvardi2009chiral,zakeri2010asymmetric}. From this point of view, low damping films based on amorphous CoFeB films or Heusler compounds hold great promise to the chiral magnonics. The symmetry of DMI results in a pair energy written $-\mathbf{D} \cdot \left(\mathbf{S}_i \times \mathbf{S}_j\right)$, where the orientation of the DMI vector $\mathbf{D}$ depends on the geometry\cite{Crepieux1998} favouring an orthogonal ordering of neighbouring spin moments $\mathbf{S}_i$ and $\mathbf{S}_j$. Its presence can give rise to exotic structures such as chiral N\'{e}el walls\cite{kubetzka2002spin}, cycloids\cite{ferriani2008atomic}, helices\cite{Uchida2006}, or skyrmions \cite{heinze2011spontaneous}, which opens up an unexplored field of magnetism. Its determination is not only important for fundamental understanding of its origin, most commonly studied by \textit{ab-initio} calculations\cite{yang2015anatomy}, but also for applications in spintronics \cite{Fert2013}. Indeed, knowledge of the DMI constant is important for designing an optimum footpath towards formation of isolated skyrmions\cite{boulle2016,hrabec2016current}, their topological stability\cite{rohart2016path} and to chiral magnonics\cite{kim2016spin,garcia2015narrow}.

\begin{figure}[h]
\includegraphics[width=8cm]{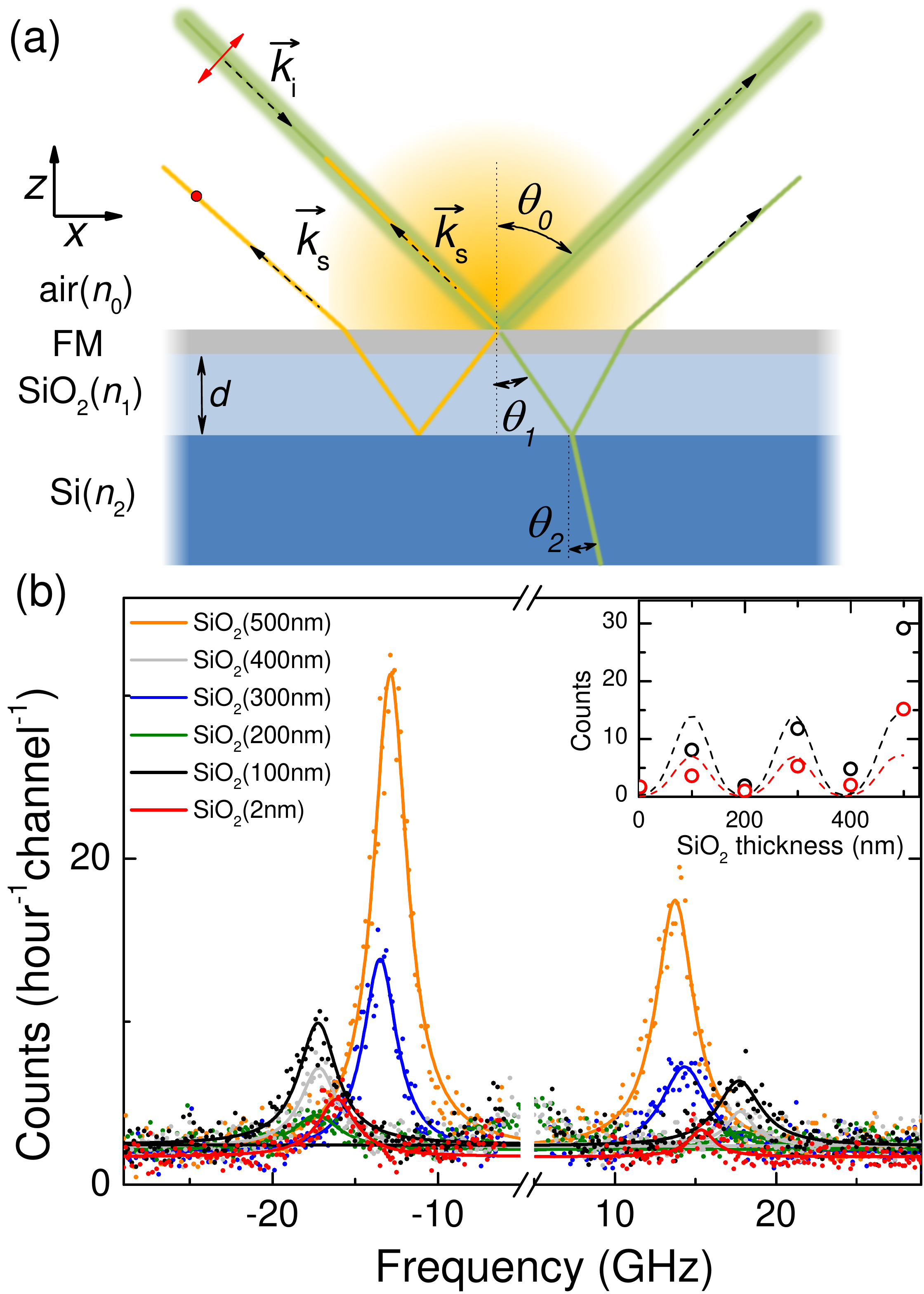}
\caption{(a) Brillouin light spectroscopy experimental setup in back-scattering geometry where the inelastically scattered light is collected in the direction of the incident $p$-polarized light.
The schematic highlights the multiple interference effects that affect the incident and scattered light. Polarization of the light is indicated in red. (b) Spectra measured at $\theta_0= 30^{\circ}$ at $0.5\,$T for SiO$_2$(500, 400, 300, 200, 100\,nm)$\backslash$Pt$\backslash$Co$\backslash$Cu and at $0.3\,$T for SiO$_2$(2\,nm)$\backslash$Cu$\backslash$Co$\backslash$Pt.
The solid lines correspond to the Lorentzian fits. Note the cut in the frequency axis. The inset shows comparison between the measured intensities at $\theta_0= 30^{\circ}$ and the calculated intensities obtained by the magnetism-sensitive method\cite{roussigne1995brillouin} for Stokes (black) and anti-Stokes modes (red). \label{Fig_1}}
\end{figure}

Several methods have been proposed to quantify the DMI in ultrathin films based on the physics of domain walls, for example by the determination of a stopping magnetic field\cite{ryu2013chiral,emori2013current,conte2015role,Choe,hrabec2014measuring}, an external magnetic field compensating the effective DMI field within a domain wall\cite{Thiaville_DMI}.
While the electric current-based experiments require complex transport measurements, the field-based methods still face the lack of full fundamental understanding\cite{vavnatka2015velocity,lavrijsen2015asymmetric,pellegren2016nonlocal}. The most robust method for measuring DMI is BLS, providing an access to the spin waves which, in the presence of DMI, show an energy dependence on the propagation sense\cite{di2015direct,belmeguenai2015interfacial,nembach2015linear}. Here the light's electric field couples to the magnetization waves propagating in antiparallel directions, producing an inelastically scattered light with a magnon footprint.

In this Letter we show that the intensity of the inelastically scattered light is strongly dependent on the thickness of the used dielectric underlayer (here SiO$_2$). We show that this effect can be simply explained by arguments of thin films optics, that are moreover in qualitative agreement with a numerical model based on the microscopic interaction of light with the magnetic modes. These models provide a guidance for the selection of the substrate for enhancing the signal at all incidence angles, typically speeding up the measurements by an order of magnitude.

Thin films of Pt(5)$\backslash$Co(1.2)$\backslash$Cu(3)$\backslash$Pt(2) and Cu(3)$\backslash$Co(1.2)$\backslash$Pt(5) (all thicknesses in nanometers; we use the convention that the bottom layer is written first) were grown in a ultra-high vacuum evaporator with base pressure of $10^{-10}$~mBar on a silicon substrate with variably thick SiO$_2$ layer. Samples were grown in three batches: 1) SiO$_2$(2~nm) 2) SiO$_2$(500~nm), SiO$_2$(300~nm) and 3) SiO$_2$(100~nm), SiO$_2$(200~nm), SiO$_2$(400~nm). The optical properties and thickness of SiO$_2$ layer have been determined by ellipsometry. Superconducting quantum interference device (SQUID) magnetometry has been used to measure hysteresis loops in order to determine the magnetization $M_\mathrm{s}$ at saturation and the anisotropy field $\mu_0H_\mathrm{K}$. The spectrometer is a J.R. Sandercock product where the distance $\delta$ between the reflecting surfaces of the Fabry-Perot interferometer is measured by means of an analog comparator micrometer.  The free spectral range $c/2\delta$ is  thus  accurately determined. During the acquisition of the spectrum, the distance  between the reflecting surfaces is controlled by the software which allows calibration of the frequency increment associated to one channel. Glass is regularly used to check if the set-up operates correctly: the positions of the Stokes and anti-Stokes lines should be the same (a shift can occur if the sample beam and the reference beam are misaligned) and should correspond to the longitudinal mode frequency of silica. We have used a crossed analyzer in order to eliminate non-magnetic modes such as phonons, as shown for example in Ref.~\onlinecite{rowan2017interfacial}. The BLS setup is employed in the Damon-Eshbach (DE) geometry where the magnetic field is applied perpendicular to the incidence plane, which allows spin waves propagating along the in-plane direction perpendicular to the applied field to be probed. In our experimental setup sketched in Fig.~\ref{Fig_1}(a) the incident light (defined by $\lambda=532$~nm and incidence angle $\theta_0$) hits the probed surface and while the light is refracted (green path), a part of the light is also inelastically scattered (orange). In order to maximize the momentum transferred, we collect the back-scattered light component $k_\mathrm{s}$ (orange path).

\begin{figure}[h]
\includegraphics[width=8cm]{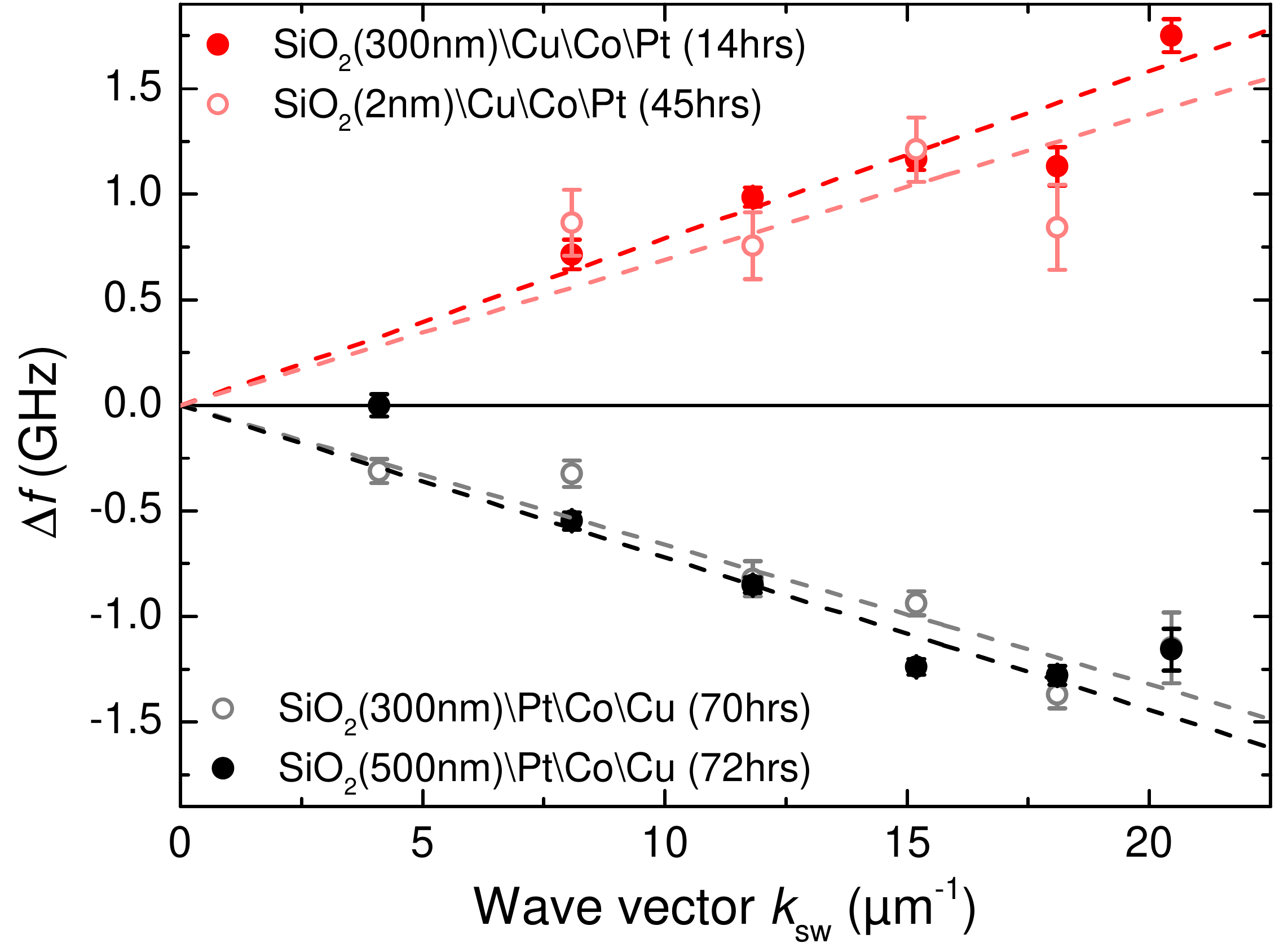}
\caption{Measured spin wave dispersion in Cu$\backslash$Co$\backslash$Pt and Pt$\backslash$Co$\backslash$Cu samples as a function of the wave vector $k_\mathrm{sw}$. The dashed lines correspond to the linear fit using equation (1). Labels show the total accumulation time for each dataset. \label{Fig_2}}
\end{figure}

Fig.~\ref{Fig_1}(b) displays experimentally measured light intensity spectra of films deposited on substrates with different SiO$_2$ thickness $d$. One can immediately see a strong variation of the signal with $d$, with vanishing intensity when only native SiO$_2$ is present. The resonance frequency at $k_\mathrm{sw}=0$ is $f_\mathrm{r}=\sqrt{f_xf_z}$, where the two principal frequencies read $f_x=(\gamma_0/2\pi)H_y$ and $f_z=(\gamma_0/2\pi)(H_y-H_\mathrm{K})$ with $H_\mathrm{K}$ being the effective anisotropy field. As shown below, these two frequencies shift linearly in $k_x$ in the presence of DMI. The position of the peaks reveals small anisotropy variations between the three sample batches. In the given geometry the spin waves propagate in the plane of incidence with $k_\mathrm{sw}= \pm 4\pi \sin(\theta_0)/\lambda$. The Stokes $f_\mathrm{S}$ (negative frequency relative to that of the incident light) and anti-Stokes $f_\mathrm{AS}$ (positive frequency) modes were then determined from Lorentzian fits to the BLS spectra. Such spectra are measured for various angles of incidence between $10^\circ$ and $70^\circ$, i.e. for various spin wave vectors $k_\mathrm{sw}$. In the presence of DMI the difference of the two frequencies reads\cite{cortes2013influence,di2015direct,belmeguenai2015interfacial}
\begin{equation}
\Delta f = f_\mathrm{S}-f_\mathrm{AS}=\frac{2\gamma}{\pi M_\mathrm{s}}D_\mathrm{eff} k_\mathrm{sw}
\end{equation}
where $D_\mathrm{eff}$ is the effective (thickness-averaged) micromagnetic DMI constant. This dispersion is illustrated in Fig.~\ref{Fig_2}, and by linear fitting using Eq.~(1) we find that $D_{\mathrm{eff}}=-0.49\pm0.04$\,mJ/m$^2$ for SiO$_2$(300~nm)$\backslash$Pt$\backslash$Co$\backslash$Cu and $D_{\mathrm{eff}}=0.56\pm0.03$\,mJ/m$^2$ for SiO$_2$(300~nm)$\backslash$Cu$\backslash$Co$\backslash$Pt respectively. Such asymmetry is expected from simple arguments of inverting the inversion symmetry by repositioning the heavy metals\cite{hrabec2016current}.

The thin-film optics theory has been already used to find the ideal thickness of Si$\backslash$SiO$_2$ for visualizing graphene, i.e. a single monolayer of carbon\cite{blake2007making}.
While in the case of graphene the calculation evaluates the interference between light multiply reflected at the various interfaces, in our case one has to consider light which is multiply reflected, inelastically scattered within the magnetic layer, the inelastic light being also multiply reflected, as indicated in Fig.~\ref{Fig_1}(a). The same arguments have also been applied to the detection of the vibration modes of graphene by Raman light scattering at normal incidence\cite{wang2008interference,yoon2009interference}. Moreover, as we need to consider inclined incidence, polarization matters and the magnetically scattered light has a polarization normal to that of the incoming light. Thus, we start with a pure thin-film optics calculation, for a simplified system Si$\backslash$SiO$_2$($d$)$\backslash$air with respective refractive indices $n_2$, $n_1$ and $n_0=1$, both Si and air of infinite thickness while the dielectric thickness is $d$. We moreover study the limit of an infinitely thin magnetic layer, thus evaluate at the SiO$_2$$\backslash$air interface the electric field at the frequency of the incident light, as well as the electric field that is inelastically backscattered.

For the first calculation, the continuity at the SiO$_2$$\backslash$air interface of the tangential component of electric and magnetic fields \cite{jackson1999classical} gives, taking into account the incident $p$ polarization [indicated by red in Fig.~\ref{Fig_1}(a)]
\begin{eqnarray*}\label{Eq_cont1}
% \nonumber % Remove numbering (before each equation)
  \left(E_0+E'_0\right)\cos\theta_0&=&\cos\theta_1 \left(E_1+E'_1\right)\\ \nonumber
  E_0-E'_0&=&n_1\left(E_1-E'_1\right),
\end{eqnarray*}
where $E_0$ is the incident electric field amplitude ($p$ polarization) in air, $E'_0$ the same for the reflected beam, the subscripts 1 and 2 applying to SiO$_2$ and Si, respectively.
At the Si$\backslash$SiO$_2$ interface one has similarly
\begin{eqnarray*}\label{Eq_cont2}
% \nonumber % Remove numbering (before each equation)
  \left(E_1e^{i \Phi}+E'_1e^{-i \Phi}\right)\cos\theta_1&=&E_2\cos\theta_2\\ \nonumber
   \left(E_1e^{i \Phi}-E'_1e^{-i \Phi}\right)n_1&=&n_2E_2,
\end{eqnarray*}
where $\Phi=2i\pi n_1\cos{\theta_1}d/\lambda$ is the phase acquired upon transmission through SiO$_2$.
Solving these equations gives the amplification factor (also called absorption factor in the Raman community\cite{yoon2009interference}) $A \equiv \frac{E_0+E'_0}{E_0}$ of the electric field at the SiO$_2$$\backslash$air interface with respect to the incident electric field
%\begin{equation}\label{Eq_ThicknDep}
%  E_1=E_1\exp\left(2ik\cos{\theta}d\right)\frac{n_1\cos_{\theta_2}-n_2\cos{\theta_1}}{n_1\cos_{\theta_2}+n_2\cos{\theta_1}}.
%\end{equation}
%\begin{widetext}
\begin{equation}\label{Eq_cont3}
A=\frac{2\cos\theta_1 \left(1+r_\mathrm{p} e^{2i\Phi}\right)}
{\left(\cos\theta_1+n_1\cos\theta_0\right)+r_\mathrm{p} e^{2i\Phi}\left(\cos\theta_1-n_1\cos\theta_0\right)},
\end{equation}
%\end{widetext}
where we used the Fresnel reflection coefficient at the Si$\backslash$SiO$_2$ interface \newline
$r_\mathrm{p}=\left(n_1\cos\theta_2-n_2\cos\theta_1\right)/\left(n_1\cos\theta_2+n_2\cos\theta_1\right)$.

\begin{figure}[h]
\includegraphics[width=8cm]{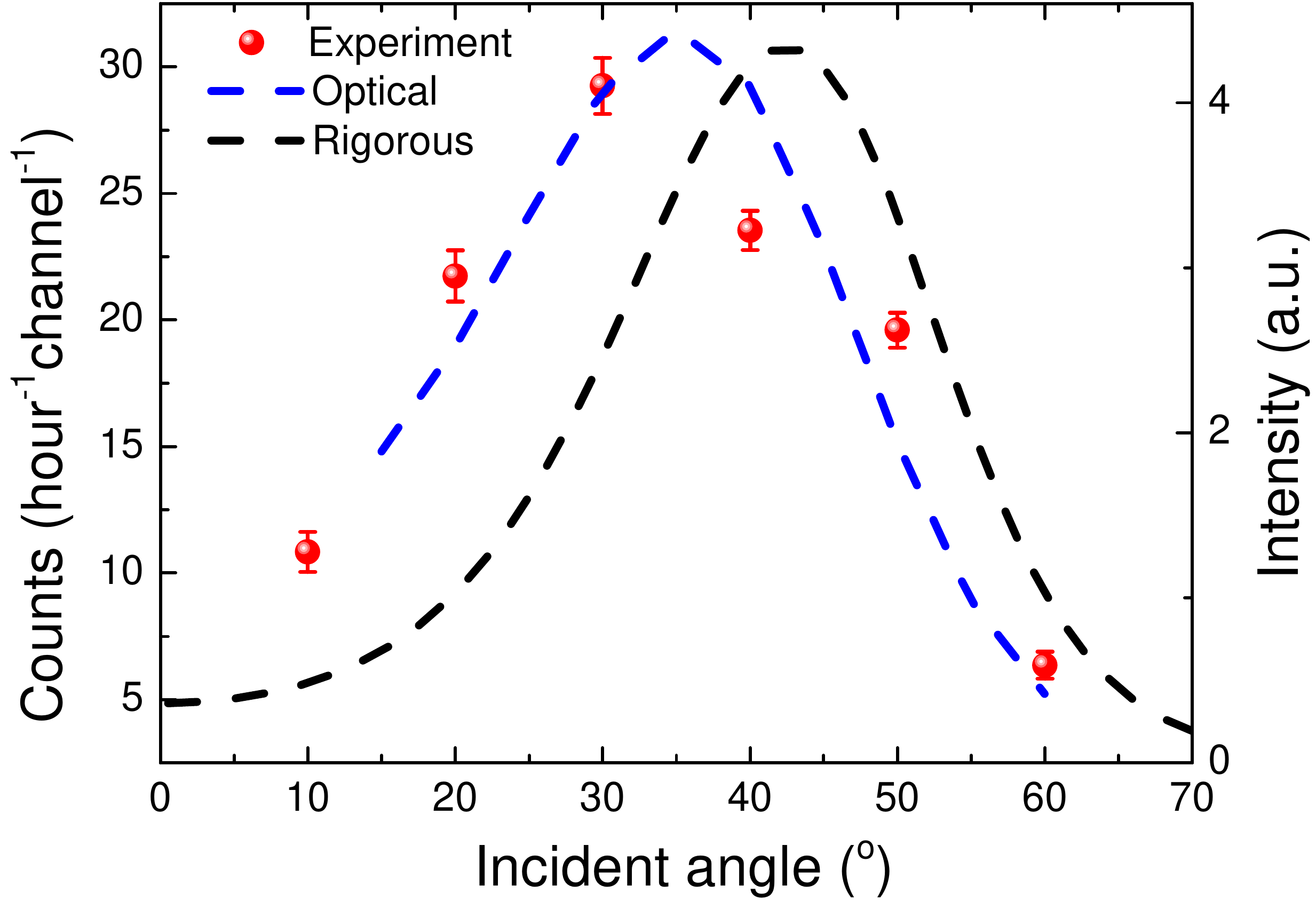}
\caption{Intensity of the Stokes peak measured for various angles at $0.5$~T in SiO$_2$(500~nm)$\backslash$Pt$\backslash$Co$\backslash$Cu sample. The fitted curves correspond to the calculation expressed by Eqs.~(\ref{Eq_cont3}, \ref{Eq_cont6}), and by the more rigorous model described in Ref.~\onlinecite{roussigne1995brillouin}. \label{Fig_3}}
\end{figure}

For the second calculation, the dielectric response of the infinitely thin magnetic layer is described by a surface current density.
The magnetically scattered light is created by this current due to the magneto-optical effect, namely
$\vec{J}_m= \omega' \epsilon_0 Q \vec{E} \times \vec{m}$, where $\omega'$ is the angular frequency of the scattered light, $Q$ the magneto-optical constant \cite{hubert1998magnetic} and $\vec{m}$ the spin-wave amplitude, a complex vector with no $y$ component as the DC magnetization is along that axis, so that here one has $\vec{J}_m = J_m \vec{y}$.
The electric field $\vec{E}$ inside the magnetic film has an in-plane component that is continuous across the interfaces, and an out of plane component that is not continuous, as continuity applies to the electric displacement $D_z=\epsilon E_z$.
Given the large refractive index of metals, and the fact that we are looking for enhanced in-plane electric field components, a reasonable approximation is to neglect the out of plane field components in computing the magneto-optical equivalent surface current $\vec{J}_m$.
In this approximation, the magnetic current is simply proportional to the first amplification factor $A$.
Using the same numbering for the fields (symbols $F$) at angular frequency $\omega'$ and with an $s$ polarization, one writes at the top interface
\begin{eqnarray*}\label{Eq_cont4}
% \nonumber % Remove numbering (before each equation)
  F'_0&=&F_1+F'_1\\ \nonumber
  F'_0 \cos\theta_0 &=& n_1 \cos\theta_1 \left(F'_1-F_1\right) + J,
\end{eqnarray*}
with $J=-c \mu_0 J_{m}$.
For the second interface on has simply
\begin{eqnarray*}\label{Eq_cont5}
% \nonumber % Remove numbering (before each equation)
  F_1e^{i \Phi}+F'_1e^{-i \Phi}&=&F_2\\ \nonumber
  n_1 \cos\theta_1 \left(F_1e^{i \Phi}-F'_1e^{-i \Phi}\right) &=&  n_2 \cos\theta_2 F_2.
\end{eqnarray*}
The amplification factor (also called scattering factor\cite{yoon2009interference}) $B \equiv\frac{F'_0}{J}$ of the backscattered light by multiple interferences then reads
%\begin{widetext}
\begin{equation}\label{Eq_cont6}
B=\frac{\left(1+r_\mathrm{s} e^{2i\Phi}\right)}
{\left(\cos\theta_0+n_1\cos\theta_1\right)+r_\mathrm{s} e^{2i\Phi}\left(\cos\theta_0-n_1\cos\theta_1\right)},
\end{equation}
%\end{widetext}
where now the other Fresnel reflection coefficient at the Si$\backslash$SiO$_2$ interface applies
$r_\mathrm{s}=\left(n_1\cos\theta_1-n_2\cos\theta_2\right)/\left(n_1\cos\theta_1+n_2\cos\theta_2\right)$.
The collected backscattered light intensity is then simply given by $|A|^2|B|^2 \cos \theta_0$, the angular factor
coming from the probed density of excitations in the evaluation of the scattering cross-section \cite{loudon1980analysis}.
The reference is the free-standing layer, where $A=B=1$.

\begin{figure}[h!]
\includegraphics[width=8cm]{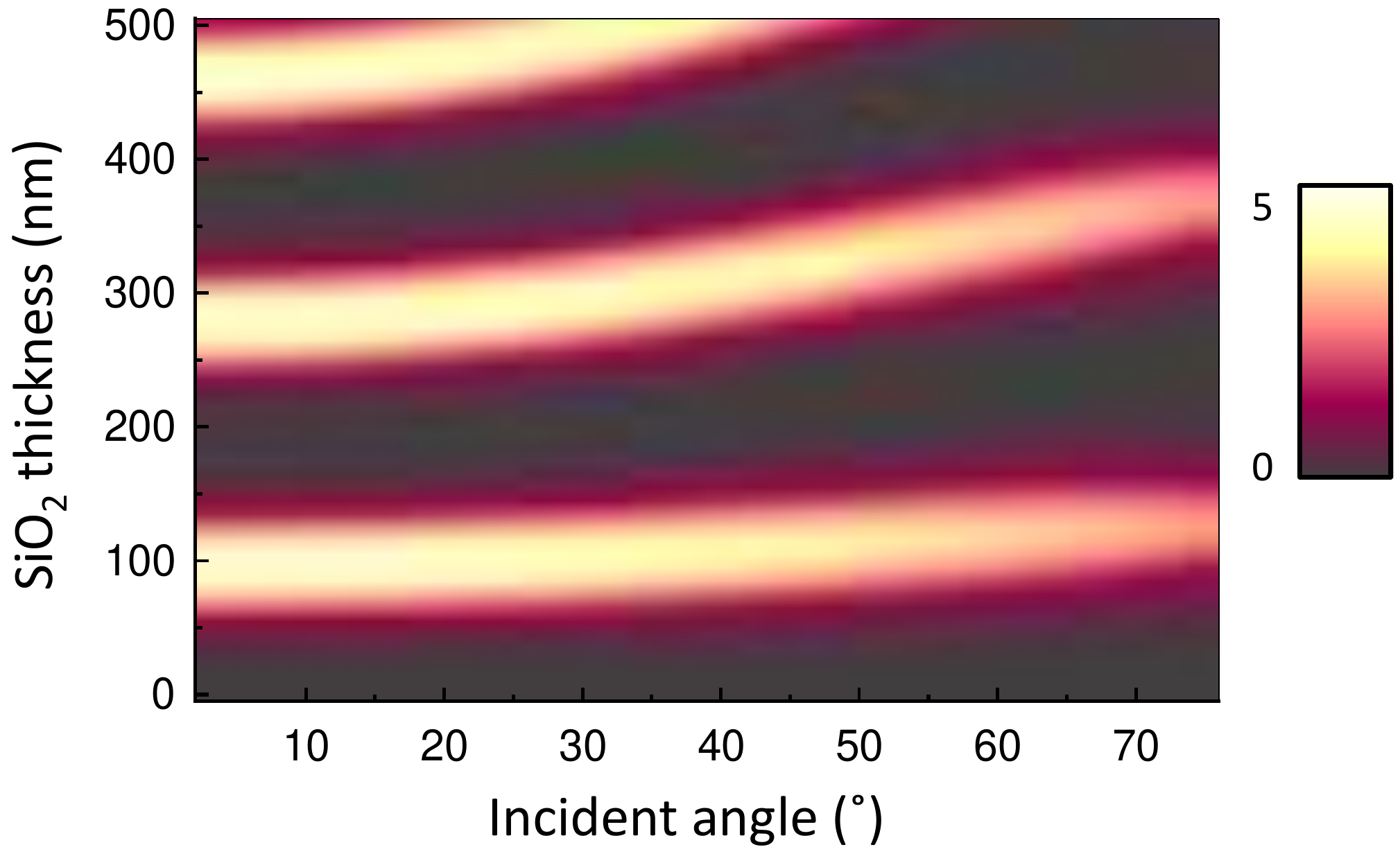}
\caption{Calculated light intensities $|A|^2|B|^2 \cos \theta_0$ for various thicknesses $d$ of SiO$_2$ using Eqs.~(\ref{Eq_cont3}, \ref{Eq_cont6}) for a simplified system  Si$\backslash$SiO$_2$$\backslash$air. The same features are obtained by the more rigorous model described in Ref.~\onlinecite{roussigne1995brillouin} for both Stokes and anti-Stokes modes. The used parameters are $n_1=1.46$, $n_2=4.142+0.032i$ and $\lambda=532$\,nm.
\label{Fig_4}}
\end{figure}

The above calculation ignores the microscopic details of the magneto-optical coupling between electric field and magnetization of the spinwaves, whereas it is known, for example, to be responsible for the S/AS intensity asymmetry \cite{stashkevich2009spin}. A full electromagnetic and micromagnetic calculation was developed earlier by one of us\cite{roussigne1995brillouin}, and was applied to the present situation.

The two models have been compared to the measured angular dependence of the peak intensity, as plotted in Fig.~\ref{Fig_3}. Since the simplified model is insensitive to the magnetization, the calculated intensity using Eqs.~(\ref{Eq_cont3}, \ref{Eq_cont6}) was renormalized. Note that similar behaviour is also observed for anti-Stokes modes (not shown). The comparison between the two models shows that the purely optical model is, for this situation, sufficiently accurate. The full angular dependence of the light intensity was also computed for the same parameters (Fig.~\ref{Fig_4}). It reveals a strong variation of backscattered intensity as a function of incidence angle, as well as its significant boost-up for certain dielectric spacer thicknesses. This is qualitatively consistent with the measured intensities shown in Fig.~\ref{Fig_1}(b) and the comparison is depicted in the inset. From Fig.~\ref{Fig_4} we therefore conclude that the optimum SiO$_2$ thickness for the DMI measurements is around 90~nm ($\approx \lambda/4 n_1$) where the intensity of the light is maximum and weakly dependent on the light incidence angle. This contrasts with the standard value of $\sim300$~nm ($\approx 3\lambda/4 n_1$) used in Raman studies, where the incidence angle is not relevant.

A measurement which is related to BLS is that of the Kerr polarization rotation. In the zero sample thickness limit, the Kerr complex angle $\theta_\mathrm{K}$ is given by
$\tan \theta_\mathrm{K}=A B / (A-1)$, thus partially but not completely related to the product $AB$ that we investigated here. It is well known \cite{hubert1998magnetic,kranz1963moglichkeiten} that the Kerr rotation angle for a bulk metallic sample can be improved by coating it with an anti-reflection layer (bringing $A$ to 1, schematically). Thus the natural questions: Why not do the same here, and what if it is done? Firstly, a bulk sample is opaque so there is no point in depositing it on an underlayer, the only possible action is to coat it. For an ultrathin sample, however, both underlayer and capping layer may be involved. Here we show that the underlayer is essential. Indeed, the bottomline is that, as the electric field is zero in a perfect conductor and the tangential electric field is continuous, an ultrathin sample in contact with a metal sees no optical electric field, so gives no magneto-optical nor BLS signal. In reality the substrate is not a perfect metal. In the normal incidence case, the field $E_\mathrm{T}$ at the substrate surface is given by $E_\mathrm{T}/E_0=2/(1+n_2)$. This shows that any substrate with a large (real or complex) refractive index attenuates the optical electric field. When an underlayer is added, in the same conditions one obtains $|E_\mathrm{T}/E_0|_\mathrm{max}=2/(1+n_1^2/n_2)$ in the case where $n_2$ is real, a good approximation for Si. This shows that the addition of an underlayer turns the high $n_2$ into an asset, with a maximum maximorum $E_\mathrm{T}/E_0=2$ obtained for a perfect mirror substrate, twice the value for a free standing sample. It is key to note that the field quenching at the surface of a high refractive index substrate remains whatever the capping. In the presence of both capping and underlayer with real refractive indices, calculation shows that the maxima of BLS intensity are the same as those with the underlayer only. This is intuitive: to have the maximum field at the sample position one just needs to have the substrate reflection in phase with the incident field. Thus, in the zero sample thickness limit, if an underlayer optimizes the BLS signal, no signal can be gained with a capping layer. As the sample thickness increases, one goes continuously to the bulk case where the capping layer also becomes important.

To conclude, we have shown both experimentally and theoretically a strong dependence of the BLS signal on the used substrate. Compared to previous work on interference enhancement of Raman contrast, we have addressed the additional requirements of $k$-dependent magnetic inelastic scattering. Our measurements suggest that using an optimum substrate can very significantly enhance the measured signal resulting into the decrease of required measuring time, and improves the data reliability which is crucial for the determination of the DMI constant $D_\mathrm{eff}$. BLS combined with benchmark substrates therefore can become a routine characterization technique for studying the DMI. Moreover, we have proposed a guideline to facilitate the magnon visualization in ultrathin magnetic films.

This work has been supported by the Agence Nationale de la Recherche under contracts ANR-14-CE26-0012 (Ultrasky) and ANR-09-NANO-002 (Hyfont), the RTRA Triangle de la Physique (Multivap), the Conseil r\'{e}gional d'{\^{I}}le-de-France through the DIM C'Nano (Imadyn) and NanoK (Bidul). We thank Marion Grzelka for help with ellipsometry measurements. We also thank Cedric Villebasse and Jean-Paul Adam for providing us some of the SiO$_2$ substrates.


\begin{thebibliography}{44}%
\makeatletter
\providecommand \@ifxundefined [1]{%
 \@ifx{#1\undefined}
}%
\providecommand \@ifnum [1]{%
 \ifnum #1\expandafter \@firstoftwo
 \else \expandafter \@secondoftwo
 \fi
}%
\providecommand \@ifx [1]{%
 \ifx #1\expandafter \@firstoftwo
 \else \expandafter \@secondoftwo
 \fi
}%
\providecommand \natexlab [1]{#1}%
\providecommand \enquote  [1]{``#1''}%
\providecommand \bibnamefont  [1]{#1}%
\providecommand \bibfnamefont [1]{#1}%
\providecommand \citenamefont [1]{#1}%
\providecommand \href@noop [0]{\@secondoftwo}%
\providecommand \href [0]{\begingroup \@sanitize@url \@href}%
\providecommand \@href[1]{\@@startlink{#1}\@@href}%
\providecommand \@@href[1]{\endgroup#1\@@endlink}%
\providecommand \@sanitize@url [0]{\catcode `\\12\catcode `\$12\catcode
  `\&12\catcode `\#12\catcode `\^12\catcode `\_12\catcode `\%12\relax}%
\providecommand \@@startlink[1]{}%
\providecommand \@@endlink[0]{}%
\providecommand \url  [0]{\begingroup\@sanitize@url \@url }%
\providecommand \@url [1]{\endgroup\@href {#1}{\urlprefix }}%
\providecommand \urlprefix  [0]{URL }%
\providecommand \Eprint [0]{\href }%
\providecommand \doibase [0]{http://dx.doi.org/}%
\providecommand \selectlanguage [0]{\@gobble}%
\providecommand \bibinfo  [0]{\@secondoftwo}%
\providecommand \bibfield  [0]{\@secondoftwo}%
\providecommand \translation [1]{[#1]}%
\providecommand \BibitemOpen [0]{}%
\providecommand \bibitemStop [0]{}%
\providecommand \bibitemNoStop [0]{.\EOS\space}%
\providecommand \EOS [0]{\spacefactor3000\relax}%
\providecommand \BibitemShut  [1]{\csname bibitem#1\endcsname}%
\let\auto@bib@innerbib\@empty
%</preamble>
\bibitem [{\citenamefont {Bloch}(1930)}]{bloch1930theorie}%
  \BibitemOpen
  \bibfield  {author} {\bibinfo {author} {\bibfnamefont {F.}~\bibnamefont
  {Bloch}},\ }\href@noop {} {\bibfield  {journal} {\bibinfo  {journal} {Z.
  Phys.}\ }\textbf {\bibinfo {volume} {61}},\ \bibinfo {pages} {206} (\bibinfo
  {year} {1930})}\BibitemShut {NoStop}%
\bibitem [{\citenamefont {Owens}\ \emph {et~al.}(1985)\citenamefont {Owens},
  \citenamefont {Collins},\ and\ \citenamefont {Carter}}]{owens1985system}%
  \BibitemOpen
  \bibfield  {author} {\bibinfo {author} {\bibfnamefont {J.}~\bibnamefont
  {Owens}}, \bibinfo {author} {\bibfnamefont {J.}~\bibnamefont {Collins}}, \
  and\ \bibinfo {author} {\bibfnamefont {R.}~\bibnamefont {Carter}},\
  }\href@noop {} {\bibfield  {journal} {\bibinfo  {journal} {Circuits, systems,
  and signal processing}\ }\textbf {\bibinfo {volume} {4}},\ \bibinfo {pages}
  {317} (\bibinfo {year} {1985})}\BibitemShut {NoStop}%
\bibitem [{\citenamefont {Wintz}\ \emph {et~al.}(2016)\citenamefont {Wintz},
  \citenamefont {Tiberkevich}, \citenamefont {Weigand}, \citenamefont {Raabe},
  \citenamefont {Lindner}, \citenamefont {Erbe}, \citenamefont {Slavin},\ and\
  \citenamefont {Fassbender}}]{wintz2016magnetic}%
  \BibitemOpen
  \bibfield  {author} {\bibinfo {author} {\bibfnamefont {S.}~\bibnamefont
  {Wintz}}, \bibinfo {author} {\bibfnamefont {V.}~\bibnamefont {Tiberkevich}},
  \bibinfo {author} {\bibfnamefont {M.}~\bibnamefont {Weigand}}, \bibinfo
  {author} {\bibfnamefont {J.}~\bibnamefont {Raabe}}, \bibinfo {author}
  {\bibfnamefont {J.}~\bibnamefont {Lindner}}, \bibinfo {author} {\bibfnamefont
  {A.}~\bibnamefont {Erbe}}, \bibinfo {author} {\bibfnamefont {A.}~\bibnamefont
  {Slavin}}, \ and\ \bibinfo {author} {\bibfnamefont {J.}~\bibnamefont
  {Fassbender}},\ }\href@noop {} {\bibfield  {journal} {\bibinfo  {journal}
  {Nat. Nanotech.}\ }\textbf {\bibinfo {volume} {11}},\ \bibinfo {pages} {948–}
  (\bibinfo {year} {2016})}\BibitemShut {NoStop}%
\bibitem [{\citenamefont {Mook}\ and\ \citenamefont
  {Paul}(1985)}]{mook1985neutron}%
  \BibitemOpen
  \bibfield  {author} {\bibinfo {author} {\bibfnamefont {H.}~\bibnamefont
  {Mook}}\ and\ \bibinfo {author} {\bibfnamefont {D.~M.}\ \bibnamefont
  {Paul}},\ }\href@noop {} {\bibfield  {journal} {\bibinfo  {journal} {Phys.
  Rev. Lett.}\ }\textbf {\bibinfo {volume} {54}},\ \bibinfo {pages} {227}
  (\bibinfo {year} {1985})}\BibitemShut {NoStop}%
\bibitem [{\citenamefont {Demokritov}\ \emph {et~al.}(2001)\citenamefont
  {Demokritov}, \citenamefont {Hillebrands},\ and\ \citenamefont
  {Slavin}}]{demokritov2001brillouin}%
  \BibitemOpen
  \bibfield  {author} {\bibinfo {author} {\bibfnamefont {S.~O.}\ \bibnamefont
  {Demokritov}}, \bibinfo {author} {\bibfnamefont {B.}~\bibnamefont
  {Hillebrands}}, \ and\ \bibinfo {author} {\bibfnamefont {A.~N.}\ \bibnamefont
  {Slavin}},\ }\href@noop {} {\bibfield  {journal} {\bibinfo  {journal} {Phys.
  Rep.}\ }\textbf {\bibinfo {volume} {348}},\ \bibinfo {pages} {441} (\bibinfo
  {year} {2001})}\BibitemShut {NoStop}%
\bibitem [{\citenamefont {Kruglyak}\ \emph {et~al.}(2010)\citenamefont
  {Kruglyak}, \citenamefont {Demokritov},\ and\ \citenamefont
  {Grundler}}]{kruglyak2010magnonics}%
  \BibitemOpen
  \bibfield  {author} {\bibinfo {author} {\bibfnamefont {V.}~\bibnamefont
  {Kruglyak}}, \bibinfo {author} {\bibfnamefont {S.}~\bibnamefont
  {Demokritov}}, \ and\ \bibinfo {author} {\bibfnamefont {D.}~\bibnamefont
  {Grundler}},\ }\href@noop {} {\bibfield  {journal} {\bibinfo  {journal} {J.
  Phys. D: Appl. Phys.}\ }\textbf {\bibinfo {volume} {43}},\ \bibinfo {pages}
  {264001} (\bibinfo {year} {2010})}\BibitemShut {NoStop}%
\bibitem [{\citenamefont {Cherepanov}\ \emph {et~al.}(1993)\citenamefont
  {Cherepanov}, \citenamefont {Kolokolov},\ and\ \citenamefont
  {L'vov}}]{cherepanov1993saga}%
  \BibitemOpen
  \bibfield  {author} {\bibinfo {author} {\bibfnamefont {V.}~\bibnamefont
  {Cherepanov}}, \bibinfo {author} {\bibfnamefont {I.}~\bibnamefont
  {Kolokolov}}, \ and\ \bibinfo {author} {\bibfnamefont {V.}~\bibnamefont
  {L'vov}},\ }\href@noop {} {\bibfield  {journal} {\bibinfo  {journal} {Phys.
  Rep.}\ }\textbf {\bibinfo {volume} {229}},\ \bibinfo {pages} {81} (\bibinfo
  {year} {1993})}\BibitemShut {NoStop}%
\bibitem [{\citenamefont {Udvardi}\ and\ \citenamefont
  {Szunyogh}(2009)}]{udvardi2009chiral}%
  \BibitemOpen
  \bibfield  {author} {\bibinfo {author} {\bibfnamefont {L.}~\bibnamefont
  {Udvardi}}\ and\ \bibinfo {author} {\bibfnamefont {L.}~\bibnamefont
  {Szunyogh}},\ }\href@noop {} {\bibfield  {journal} {\bibinfo  {journal}
  {Phys. Rev. Lett.}\ }\textbf {\bibinfo {volume} {102}},\ \bibinfo {pages}
  {207204} (\bibinfo {year} {2009})}\BibitemShut {NoStop}%
\bibitem [{\citenamefont {Zakeri}\ \emph {et~al.}(2010)\citenamefont {Zakeri},
  \citenamefont {Zhang}, \citenamefont {Prokop}, \citenamefont {Chuang},
  \citenamefont {Sakr}, \citenamefont {Tang},\ and\ \citenamefont
  {Kirschner}}]{zakeri2010asymmetric}%
  \BibitemOpen
  \bibfield  {author} {\bibinfo {author} {\bibfnamefont {K.}~\bibnamefont
  {Zakeri}}, \bibinfo {author} {\bibfnamefont {Y.}~\bibnamefont {Zhang}},
  \bibinfo {author} {\bibfnamefont {J.}~\bibnamefont {Prokop}}, \bibinfo
  {author} {\bibfnamefont {T.-H.}\ \bibnamefont {Chuang}}, \bibinfo {author}
  {\bibfnamefont {N.}~\bibnamefont {Sakr}}, \bibinfo {author} {\bibfnamefont
  {W.}~\bibnamefont {Tang}}, \ and\ \bibinfo {author} {\bibfnamefont
  {J.}~\bibnamefont {Kirschner}},\ }\href@noop {} {\bibfield  {journal}
  {\bibinfo  {journal} {Phys. Rev. Lett.}\ }\textbf {\bibinfo {volume} {104}},\
  \bibinfo {pages} {137203} (\bibinfo {year} {2010})}\BibitemShut {NoStop}%
\bibitem [{\citenamefont {Cr\'{e}pieux}\ and\ \citenamefont
  {Lacroix}(1998)}]{Crepieux1998}%
  \BibitemOpen
  \bibfield  {author} {\bibinfo {author} {\bibfnamefont {A.}~\bibnamefont
  {Cr\'{e}pieux}}\ and\ \bibinfo {author} {\bibfnamefont {C.}~\bibnamefont
  {Lacroix}},\ }\href@noop {} {\bibfield  {journal} {\bibinfo  {journal} {J.
  Magn. Magn. Mat.}\ }\textbf {\bibinfo {volume} {182}},\ \bibinfo {pages}
  {341} (\bibinfo {year} {1998})}\BibitemShut {NoStop}%
\bibitem [{\citenamefont {Kubetzka}\ \emph {et~al.}(2002)\citenamefont
  {Kubetzka}, \citenamefont {Bode}, \citenamefont {Pietzsch},\ and\
  \citenamefont {Wiesendanger}}]{kubetzka2002spin}%
  \BibitemOpen
  \bibfield  {author} {\bibinfo {author} {\bibfnamefont {A.}~\bibnamefont
  {Kubetzka}}, \bibinfo {author} {\bibfnamefont {M.}~\bibnamefont {Bode}},
  \bibinfo {author} {\bibfnamefont {O.}~\bibnamefont {Pietzsch}}, \ and\
  \bibinfo {author} {\bibfnamefont {R.}~\bibnamefont {Wiesendanger}},\
  }\href@noop {} {\bibfield  {journal} {\bibinfo  {journal} {Phys. Rev. Lett.}\
  }\textbf {\bibinfo {volume} {88}},\ \bibinfo {pages} {057201} (\bibinfo
  {year} {2002})}\BibitemShut {NoStop}%
\bibitem [{\citenamefont {Ferriani}\ \emph {et~al.}(2008)\citenamefont
  {Ferriani}, \citenamefont {von Bergmann}, \citenamefont {Vedmedenko},
  \citenamefont {Heinze}, \citenamefont {Bode}, \citenamefont {Heide},
  \citenamefont {Bihlmayer}, \citenamefont {Bl\"{u}gel},\ and\ \citenamefont
  {Wiesendanger}}]{ferriani2008atomic}%
  \BibitemOpen
  \bibfield  {author} {\bibinfo {author} {\bibfnamefont {P.}~\bibnamefont
  {Ferriani}}, \bibinfo {author} {\bibfnamefont {K.}~\bibnamefont {von
  Bergmann}}, \bibinfo {author} {\bibfnamefont {E.~Y.}\ \bibnamefont
  {Vedmedenko}}, \bibinfo {author} {\bibfnamefont {S.}~\bibnamefont {Heinze}},
  \bibinfo {author} {\bibfnamefont {M.}~\bibnamefont {Bode}}, \bibinfo {author}
  {\bibfnamefont {M.}~\bibnamefont {Heide}}, \bibinfo {author} {\bibfnamefont
  {G.}~\bibnamefont {Bihlmayer}}, \bibinfo {author} {\bibfnamefont
  {S.}~\bibnamefont {Bl\"{u}gel}}, \ and\ \bibinfo {author} {\bibfnamefont
  {R.}~\bibnamefont {Wiesendanger}},\ }\href@noop {} {\bibfield  {journal}
  {\bibinfo  {journal} {Phys. Rev. Lett.}\ }\textbf {\bibinfo {volume} {101}},\
  \bibinfo {pages} {027201} (\bibinfo {year} {2008})}\BibitemShut {NoStop}%
\bibitem [{\citenamefont {Uchida}\ \emph {et~al.}(2006)\citenamefont {Uchida},
  \citenamefont {Onose}, \citenamefont {Matsui},\ and\ \citenamefont
  {Tokura}}]{Uchida2006}%
  \BibitemOpen
  \bibfield  {author} {\bibinfo {author} {\bibfnamefont {M.}~\bibnamefont
  {Uchida}}, \bibinfo {author} {\bibfnamefont {Y.}~\bibnamefont {Onose}},
  \bibinfo {author} {\bibfnamefont {Y.}~\bibnamefont {Matsui}}, \ and\ \bibinfo
  {author} {\bibfnamefont {Y.}~\bibnamefont {Tokura}},\ }\href@noop {}
  {\bibfield  {journal} {\bibinfo  {journal} {Science}\ }\textbf {\bibinfo
  {volume} {311}},\ \bibinfo {pages} {359} (\bibinfo {year}
  {2006})}\BibitemShut {NoStop}%
\bibitem [{\citenamefont {Heinze}\ \emph {et~al.}(2011)\citenamefont {Heinze},
  \citenamefont {von Bergmann}, \citenamefont {Menzel}, \citenamefont {Brede},
  \citenamefont {Kubetzka}, \citenamefont {Wiesendanger}, \citenamefont
  {Bihlmayer},\ and\ \citenamefont {Bl{\"u}gel}}]{heinze2011spontaneous}%
  \BibitemOpen
  \bibfield  {author} {\bibinfo {author} {\bibfnamefont {S.}~\bibnamefont
  {Heinze}}, \bibinfo {author} {\bibfnamefont {K.}~\bibnamefont {von
  Bergmann}}, \bibinfo {author} {\bibfnamefont {M.}~\bibnamefont {Menzel}},
  \bibinfo {author} {\bibfnamefont {J.}~\bibnamefont {Brede}}, \bibinfo
  {author} {\bibfnamefont {A.}~\bibnamefont {Kubetzka}}, \bibinfo {author}
  {\bibfnamefont {R.}~\bibnamefont {Wiesendanger}}, \bibinfo {author}
  {\bibfnamefont {G.}~\bibnamefont {Bihlmayer}}, \ and\ \bibinfo {author}
  {\bibfnamefont {S.}~\bibnamefont {Bl{\"u}gel}},\ }\href@noop {} {\bibfield
  {journal} {\bibinfo  {journal} {Nat. Phys.}\ }\textbf {\bibinfo {volume}
  {7}},\ \bibinfo {pages} {713} (\bibinfo {year} {2011})}\BibitemShut {NoStop}%
\bibitem [{\citenamefont {Yang}\ \emph {et~al.}(2015)\citenamefont {Yang},
  \citenamefont {Thiaville}, \citenamefont {Rohart}, \citenamefont {Fert},\
  and\ \citenamefont {Chshiev}}]{yang2015anatomy}%
  \BibitemOpen
  \bibfield  {author} {\bibinfo {author} {\bibfnamefont {H.}~\bibnamefont
  {Yang}}, \bibinfo {author} {\bibfnamefont {A.}~\bibnamefont {Thiaville}},
  \bibinfo {author} {\bibfnamefont {S.}~\bibnamefont {Rohart}}, \bibinfo
  {author} {\bibfnamefont {A.}~\bibnamefont {Fert}}, \ and\ \bibinfo {author}
  {\bibfnamefont {M.}~\bibnamefont {Chshiev}},\ }\href@noop {} {\bibfield
  {journal} {\bibinfo  {journal} {Phys. Rev. Lett.}\ }\textbf {\bibinfo
  {volume} {115}},\ \bibinfo {pages} {267210} (\bibinfo {year}
  {2015})}\BibitemShut {NoStop}%
\bibitem [{\citenamefont {Fert}\ \emph {et~al.}(2013)\citenamefont {Fert},
  \citenamefont {Cros},\ and\ \citenamefont {Sampaio}}]{Fert2013}%
  \BibitemOpen
  \bibfield  {author} {\bibinfo {author} {\bibfnamefont {A.}~\bibnamefont
  {Fert}}, \bibinfo {author} {\bibfnamefont {V.}~\bibnamefont {Cros}}, \ and\
  \bibinfo {author} {\bibfnamefont {J.}~\bibnamefont {Sampaio}},\ }\href@noop
  {} {\bibfield  {journal} {\bibinfo  {journal} {Nat. Nanotech.}\ }\textbf
  {\bibinfo {volume} {8}},\ \bibinfo {pages} {152} (\bibinfo {year}
  {2013})}\BibitemShut {NoStop}%
\bibitem [{\citenamefont {Boulle}\ \emph {et~al.}(2016)\citenamefont {Boulle},
  \citenamefont {Vogel}, \citenamefont {Yang}, \citenamefont {Pizzini},
  \citenamefont {de~Souza~Chaves}, \citenamefont {Locatelli}, \citenamefont
  {Mentes}, \citenamefont {Sala}, \citenamefont {Buda-Prejbeanu}, \citenamefont
  {Klein}, \citenamefont {Belmeguenai}, \citenamefont {Roussign\'{e}},
  \citenamefont {Stashkevich}, \citenamefont {Ch{\'e}rif}, \citenamefont
  {Aballe}, \citenamefont {Foerster}, \citenamefont {Chshiev}, \citenamefont
  {Auffret}, \citenamefont {Mihai},\ and\ \citenamefont {Gilles}}]{boulle2016}%
  \BibitemOpen
  \bibfield  {author} {\bibinfo {author} {\bibfnamefont {O.}~\bibnamefont
  {Boulle}}, \bibinfo {author} {\bibfnamefont {J.}~\bibnamefont {Vogel}},
  \bibinfo {author} {\bibfnamefont {H.}~\bibnamefont {Yang}}, \bibinfo {author}
  {\bibfnamefont {S.}~\bibnamefont {Pizzini}}, \bibinfo {author} {\bibfnamefont
  {D.}~\bibnamefont {de~Souza~Chaves}}, \bibinfo {author} {\bibfnamefont
  {A.}~\bibnamefont {Locatelli}}, \bibinfo {author} {\bibfnamefont {T.~O.}\
  \bibnamefont {Mentes}}, \bibinfo {author} {\bibfnamefont {A.}~\bibnamefont
  {Sala}}, \bibinfo {author} {\bibfnamefont {L.~D.}\ \bibnamefont
  {Buda-Prejbeanu}}, \bibinfo {author} {\bibfnamefont {O.}~\bibnamefont
  {Klein}}, \bibinfo {author} {\bibfnamefont {M.}~\bibnamefont {Belmeguenai}},
  \bibinfo {author} {\bibfnamefont {Y.}~\bibnamefont {Roussign\'{e}}}, \bibinfo
  {author} {\bibfnamefont {A.}~\bibnamefont {Stashkevich}}, \bibinfo {author}
  {\bibfnamefont {S.~M.}\ \bibnamefont {Ch{\'e}rif}}, \bibinfo {author}
  {\bibfnamefont {L.}~\bibnamefont {Aballe}}, \bibinfo {author} {\bibfnamefont
  {M.}~\bibnamefont {Foerster}}, \bibinfo {author} {\bibfnamefont
  {M.}~\bibnamefont {Chshiev}}, \bibinfo {author} {\bibfnamefont
  {S.}~\bibnamefont {Auffret}}, \bibinfo {author} {\bibfnamefont {M.~I.}\
  \bibnamefont {Mihai}}, \ and\ \bibinfo {author} {\bibfnamefont
  {G.}~\bibnamefont {Gilles}},\ }\href@noop {} {\bibfield  {journal} {\bibinfo
  {journal} {Nat. Nanotech.}\ }\textbf {\bibinfo {volume} {11}},\ \bibinfo
  {pages} {449} (\bibinfo {year} {2016})}\BibitemShut {NoStop}%
\bibitem [{\citenamefont {Hrabec}\ \emph {et~al.}(2016)\citenamefont {Hrabec},
  \citenamefont {Sampaio}, \citenamefont {Belmeguenai}, \citenamefont {Gross},
  \citenamefont {Weil}, \citenamefont {Ch{\'e}rif}, \citenamefont
  {Stachkevitch}, \citenamefont {Jacques}, \citenamefont {Thiaville},\ and\
  \citenamefont {Rohart}}]{hrabec2016current}%
  \BibitemOpen
  \bibfield  {author} {\bibinfo {author} {\bibfnamefont {A.}~\bibnamefont
  {Hrabec}}, \bibinfo {author} {\bibfnamefont {J.}~\bibnamefont {Sampaio}},
  \bibinfo {author} {\bibfnamefont {M.}~\bibnamefont {Belmeguenai}}, \bibinfo
  {author} {\bibfnamefont {I.}~\bibnamefont {Gross}}, \bibinfo {author}
  {\bibfnamefont {R.}~\bibnamefont {Weil}}, \bibinfo {author} {\bibfnamefont
  {S.~M.}\ \bibnamefont {Ch{\'e}rif}}, \bibinfo {author} {\bibfnamefont
  {A.}~\bibnamefont {Stachkevitch}}, \bibinfo {author} {\bibfnamefont
  {V.}~\bibnamefont {Jacques}}, \bibinfo {author} {\bibfnamefont
  {A.}~\bibnamefont {Thiaville}}, \ and\ \bibinfo {author} {\bibfnamefont
  {S.}~\bibnamefont {Rohart}},\ }\href@noop {} {\bibfield  {journal} {\bibinfo
  {journal} {Nat. Commun.}\ } \textbf {\bibinfo {volume} {8}},\ \bibinfo
  {pages} {15765} (\bibinfo {year}
  {2017})}\BibitemShut {NoStop}%
\bibitem [{\citenamefont {Rohart}\ \emph {et~al.}(2016)\citenamefont {Rohart},
  \citenamefont {Miltat},\ and\ \citenamefont {Thiaville}}]{rohart2016path}%
  \BibitemOpen
  \bibfield  {author} {\bibinfo {author} {\bibfnamefont {S.}~\bibnamefont
  {Rohart}}, \bibinfo {author} {\bibfnamefont {J.}~\bibnamefont {Miltat}}, \
  and\ \bibinfo {author} {\bibfnamefont {A.}~\bibnamefont {Thiaville}},\
  }\href@noop {} {\bibfield  {journal} {\bibinfo  {journal} {Phys. Rev. B}\
  }\textbf {\bibinfo {volume} {93}},\ \bibinfo {pages} {214412} (\bibinfo
  {year} {2016})}\BibitemShut {NoStop}%
\bibitem [{\citenamefont {Kim}\ \emph {et~al.}(2016)\citenamefont {Kim},
  \citenamefont {Stamps},\ and\ \citenamefont {Camley}}]{kim2016spin}%
  \BibitemOpen
  \bibfield  {author} {\bibinfo {author} {\bibfnamefont {J.-V.}\ \bibnamefont
  {Kim}}, \bibinfo {author} {\bibfnamefont {R.~L.}\ \bibnamefont {Stamps}}, \
  and\ \bibinfo {author} {\bibfnamefont {R.~E.}\ \bibnamefont {Camley}},\
  }\href@noop {} {\bibfield  {journal} {\bibinfo  {journal} {Phys. Rev. Lett.}\
  }\textbf {\bibinfo {volume} {117}},\ \bibinfo {pages} {197204} (\bibinfo
  {year} {2016})}\BibitemShut {NoStop}%
\bibitem [{\citenamefont {Garcia-Sanchez}\ \emph {et~al.}(2015)\citenamefont
  {Garcia-Sanchez}, \citenamefont {Borys}, \citenamefont {Soucaille},
  \citenamefont {Adam}, \citenamefont {Stamps},\ and\ \citenamefont
  {Kim}}]{garcia2015narrow}%
  \BibitemOpen
  \bibfield  {author} {\bibinfo {author} {\bibfnamefont {F.}~\bibnamefont
  {Garcia-Sanchez}}, \bibinfo {author} {\bibfnamefont {P.}~\bibnamefont
  {Borys}}, \bibinfo {author} {\bibfnamefont {R.}~\bibnamefont {Soucaille}},
  \bibinfo {author} {\bibfnamefont {J.-P.}\ \bibnamefont {Adam}}, \bibinfo
  {author} {\bibfnamefont {R.~L.}\ \bibnamefont {Stamps}}, \ and\ \bibinfo
  {author} {\bibfnamefont {J.-V.}\ \bibnamefont {Kim}},\ }\href@noop {}
  {\bibfield  {journal} {\bibinfo  {journal} {Phys. Rev. Lett.}\ }\textbf
  {\bibinfo {volume} {114}},\ \bibinfo {pages} {247206} (\bibinfo {year}
  {2015})}\BibitemShut {NoStop}%
\bibitem [{\citenamefont {Roussign{\'e}}\ \emph {et~al.}(1995)\citenamefont
  {Roussign{\'e}}, \citenamefont {Ganot}, \citenamefont {Dugautier},
  \citenamefont {Moch},\ and\ \citenamefont {Renard}}]{roussigne1995brillouin}%
  \BibitemOpen
  \bibfield  {author} {\bibinfo {author} {\bibfnamefont {Y.}~\bibnamefont
  {Roussign{\'e}}}, \bibinfo {author} {\bibfnamefont {F.}~\bibnamefont
  {Ganot}}, \bibinfo {author} {\bibfnamefont {C.}~\bibnamefont {Dugautier}},
  \bibinfo {author} {\bibfnamefont {P.}~\bibnamefont {Moch}}, \ and\ \bibinfo
  {author} {\bibfnamefont {D.}~\bibnamefont {Renard}},\ }\href@noop {}
  {\bibfield  {journal} {\bibinfo  {journal} {Phys. Rev. B}\ }\textbf {\bibinfo
  {volume} {52}},\ \bibinfo {pages} {350} (\bibinfo {year} {1995})}\BibitemShut
  {NoStop}%
\bibitem [{\citenamefont {Ryu}\ \emph {et~al.}(2013)\citenamefont {Ryu},
  \citenamefont {Thomas}, \citenamefont {Yang},\ and\ \citenamefont
  {Parkin}}]{ryu2013chiral}%
  \BibitemOpen
  \bibfield  {author} {\bibinfo {author} {\bibfnamefont {K.-S.}\ \bibnamefont
  {Ryu}}, \bibinfo {author} {\bibfnamefont {L.}~\bibnamefont {Thomas}},
  \bibinfo {author} {\bibfnamefont {S.-H.}\ \bibnamefont {Yang}}, \ and\
  \bibinfo {author} {\bibfnamefont {S.~S.~P.}\ \bibnamefont {Parkin}},\
  }\href@noop {} {\bibfield  {journal} {\bibinfo  {journal} {Nat. Nanotech.}\
  }\textbf {\bibinfo {volume} {8}},\ \bibinfo {pages} {527} (\bibinfo {year}
  {2013})}\BibitemShut {NoStop}%
\bibitem [{\citenamefont {Emori}\ \emph {et~al.}(2013)\citenamefont {Emori},
  \citenamefont {Bauer}, \citenamefont {Ahn}, \citenamefont {Martinez},\ and\
  \citenamefont {Beach}}]{emori2013current}%
  \BibitemOpen
  \bibfield  {author} {\bibinfo {author} {\bibfnamefont {S.}~\bibnamefont
  {Emori}}, \bibinfo {author} {\bibfnamefont {U.}~\bibnamefont {Bauer}},
  \bibinfo {author} {\bibfnamefont {S.-M.}\ \bibnamefont {Ahn}}, \bibinfo
  {author} {\bibfnamefont {E.}~\bibnamefont {Martinez}}, \ and\ \bibinfo
  {author} {\bibfnamefont {G.}~\bibnamefont {Beach}},\ }\href@noop {}
  {\bibfield  {journal} {\bibinfo  {journal} {Nat. Mater.}\ }\textbf {\bibinfo
  {volume} {12}},\ \bibinfo {pages} {611} (\bibinfo {year} {2013})}\BibitemShut
  {NoStop}%
\bibitem [{\citenamefont {Lo~Conte}\ \emph {et~al.}(2015)\citenamefont
  {Lo~Conte}, \citenamefont {Martinez}, \citenamefont {Hrabec}, \citenamefont
  {Lamperti}, \citenamefont {Schulz}, \citenamefont {Nasi}, \citenamefont
  {Lazzarini}, \citenamefont {Mantovan}, \citenamefont {Maccherozzi},
  \citenamefont {Dhesi} \emph {et~al.}}]{conte2015role}%
  \BibitemOpen
  \bibfield  {author} {\bibinfo {author} {\bibfnamefont {R.}~\bibnamefont
  {Lo~Conte}}, \bibinfo {author} {\bibfnamefont {E.}~\bibnamefont {Martinez}},
  \bibinfo {author} {\bibfnamefont {A.}~\bibnamefont {Hrabec}}, \bibinfo
  {author} {\bibfnamefont {A.}~\bibnamefont {Lamperti}}, \bibinfo {author}
  {\bibfnamefont {T.}~\bibnamefont {Schulz}}, \bibinfo {author} {\bibfnamefont
  {L.}~\bibnamefont {Nasi}}, \bibinfo {author} {\bibfnamefont {L.}~\bibnamefont
  {Lazzarini}}, \bibinfo {author} {\bibfnamefont {R.}~\bibnamefont {Mantovan}},
  \bibinfo {author} {\bibfnamefont {F.}~\bibnamefont {Maccherozzi}}, \bibinfo
  {author} {\bibfnamefont {S.}~\bibnamefont {Dhesi}},  \emph {et~al.},\
  }\href@noop {} {\bibfield  {journal} {\bibinfo  {journal} {Phys. Rev. B}\
  }\textbf {\bibinfo {volume} {91}},\ \bibinfo {pages} {014433} (\bibinfo
  {year} {2015})}\BibitemShut {NoStop}%
\bibitem [{\citenamefont {Je}\ \emph {et~al.}(2013)\citenamefont {Je},
  \citenamefont {Kim}, \citenamefont {Yoo}, \citenamefont {Min}, \citenamefont
  {Lee},\ and\ \citenamefont {Choe}}]{Choe}%
  \BibitemOpen
  \bibfield  {author} {\bibinfo {author} {\bibfnamefont {S.-G.}\ \bibnamefont
  {Je}}, \bibinfo {author} {\bibfnamefont {D.-H.}\ \bibnamefont {Kim}},
  \bibinfo {author} {\bibfnamefont {S.-C.}\ \bibnamefont {Yoo}}, \bibinfo
  {author} {\bibfnamefont {B.-C.}\ \bibnamefont {Min}}, \bibinfo {author}
  {\bibfnamefont {K.-J.}\ \bibnamefont {Lee}}, \ and\ \bibinfo {author}
  {\bibfnamefont {S.-B.}\ \bibnamefont {Choe}},\ }\href {\doibase
  10.1103/PhysRevB.88.214401} {\bibfield  {journal} {\bibinfo  {journal} {Phys.
  Rev. B}\ }\textbf {\bibinfo {volume} {88}},\ \bibinfo {pages} {214401}
  (\bibinfo {year} {2013})}\BibitemShut {NoStop}%
\bibitem [{\citenamefont {Hrabec}\ \emph {et~al.}(2014)\citenamefont {Hrabec},
  \citenamefont {Porter}, \citenamefont {Wells}, \citenamefont {Benitez},
  \citenamefont {Burnell}, \citenamefont {McVitie}, \citenamefont {McGrouther},
  \citenamefont {Moore},\ and\ \citenamefont {Marrows}}]{hrabec2014measuring}%
  \BibitemOpen
  \bibfield  {author} {\bibinfo {author} {\bibfnamefont {A.}~\bibnamefont
  {Hrabec}}, \bibinfo {author} {\bibfnamefont {N.}~\bibnamefont {Porter}},
  \bibinfo {author} {\bibfnamefont {A.}~\bibnamefont {Wells}}, \bibinfo
  {author} {\bibfnamefont {M.}~\bibnamefont {Benitez}}, \bibinfo {author}
  {\bibfnamefont {G.}~\bibnamefont {Burnell}}, \bibinfo {author} {\bibfnamefont
  {S.}~\bibnamefont {McVitie}}, \bibinfo {author} {\bibfnamefont
  {D.}~\bibnamefont {McGrouther}}, \bibinfo {author} {\bibfnamefont
  {T.}~\bibnamefont {Moore}}, \ and\ \bibinfo {author} {\bibfnamefont
  {C.}~\bibnamefont {Marrows}},\ }\href@noop {} {\bibfield  {journal} {\bibinfo
   {journal} {Phys. Rev. B}\ }\textbf {\bibinfo {volume} {90}},\ \bibinfo
  {pages} {020402} (\bibinfo {year} {2014})}\BibitemShut {NoStop}%
\bibitem [{\citenamefont {Thiaville}\ \emph {et~al.}(2012)\citenamefont
  {Thiaville}, \citenamefont {Rohart}, \citenamefont {Ju\'{e}}, \citenamefont
  {Cros},\ and\ \citenamefont {Fert}}]{Thiaville_DMI}%
  \BibitemOpen
  \bibfield  {author} {\bibinfo {author} {\bibfnamefont {A.}~\bibnamefont
  {Thiaville}}, \bibinfo {author} {\bibfnamefont {S.}~\bibnamefont {Rohart}},
  \bibinfo {author} {\bibfnamefont {E.}~\bibnamefont {Ju\'{e}}}, \bibinfo
  {author} {\bibfnamefont {V.}~\bibnamefont {Cros}}, \ and\ \bibinfo {author}
  {\bibfnamefont {A.}~\bibnamefont {Fert}},\ }\href {\doibase
  10.1209/0295-5075/100/57002} {\bibfield  {journal} {\bibinfo  {journal}
  {Europhys. Lett.}\ }\textbf {\bibinfo {volume} {100}},\ \bibinfo {eid}
  {57002} (\bibinfo {year} {2012})}\BibitemShut {NoStop}%
\bibitem [{\citenamefont {Va{\v{n}}atka}\ \emph {et~al.}(2015)\citenamefont
  {Va{\v{n}}atka}, \citenamefont {Rojas-S{\'a}nchez}, \citenamefont {Vogel},
  \citenamefont {Bonfim}, \citenamefont {Belmeguenai}, \citenamefont
  {Roussign{\'e}}, \citenamefont {Stashkevich}, \citenamefont {Thiaville},\
  and\ \citenamefont {Pizzini}}]{vavnatka2015velocity}%
  \BibitemOpen
  \bibfield  {author} {\bibinfo {author} {\bibfnamefont {M.}~\bibnamefont
  {Va{\v{n}}atka}}, \bibinfo {author} {\bibfnamefont {J.-C.}\ \bibnamefont
  {Rojas-S{\'a}nchez}}, \bibinfo {author} {\bibfnamefont {J.}~\bibnamefont
  {Vogel}}, \bibinfo {author} {\bibfnamefont {M.}~\bibnamefont {Bonfim}},
  \bibinfo {author} {\bibfnamefont {M.}~\bibnamefont {Belmeguenai}}, \bibinfo
  {author} {\bibfnamefont {Y.}~\bibnamefont {Roussign{\'e}}}, \bibinfo {author}
  {\bibfnamefont {A.}~\bibnamefont {Stashkevich}}, \bibinfo {author}
  {\bibfnamefont {A.}~\bibnamefont {Thiaville}}, \ and\ \bibinfo {author}
  {\bibfnamefont {S.}~\bibnamefont {Pizzini}},\ }\href@noop {} {\bibfield
  {journal} {\bibinfo  {journal} {J. Phys. Condens. Matter.}\ }\textbf
  {\bibinfo {volume} {27}},\ \bibinfo {pages} {326002} (\bibinfo {year}
  {2015})}\BibitemShut {NoStop}%
\bibitem [{\citenamefont {Lavrijsen}\ \emph {et~al.}(2015)\citenamefont
  {Lavrijsen}, \citenamefont {Hartmann}, \citenamefont {van~den Brink},
  \citenamefont {Yin}, \citenamefont {Barcones}, \citenamefont {Duine},
  \citenamefont {Verheijen}, \citenamefont {Swagten},\ and\ \citenamefont
  {Koopmans}}]{lavrijsen2015asymmetric}%
  \BibitemOpen
  \bibfield  {author} {\bibinfo {author} {\bibfnamefont {R.}~\bibnamefont
  {Lavrijsen}}, \bibinfo {author} {\bibfnamefont {D.}~\bibnamefont {Hartmann}},
  \bibinfo {author} {\bibfnamefont {A.}~\bibnamefont {van~den Brink}}, \bibinfo
  {author} {\bibfnamefont {Y.}~\bibnamefont {Yin}}, \bibinfo {author}
  {\bibfnamefont {B.}~\bibnamefont {Barcones}}, \bibinfo {author}
  {\bibfnamefont {R.}~\bibnamefont {Duine}}, \bibinfo {author} {\bibfnamefont
  {M.}~\bibnamefont {Verheijen}}, \bibinfo {author} {\bibfnamefont
  {H.}~\bibnamefont {Swagten}}, \ and\ \bibinfo {author} {\bibfnamefont
  {B.}~\bibnamefont {Koopmans}},\ }\href@noop {} {\bibfield  {journal}
  {\bibinfo  {journal} {Phys. Rev. B}\ }\textbf {\bibinfo {volume} {91}},\
  \bibinfo {pages} {104414} (\bibinfo {year} {2015})}\BibitemShut {NoStop}%
\bibitem [{\citenamefont {Pellegren}\ \emph {et~al.}(2016)\citenamefont
  {Pellegren}, \citenamefont {Lau},\ and\ \citenamefont
  {Sokalski}}]{pellegren2016nonlocal}%
  \BibitemOpen
  \bibfield  {author} {\bibinfo {author} {\bibfnamefont {P.}~\bibnamefont
  {Pellegren}}, \bibinfo {author} {\bibfnamefont {D.}~\bibnamefont {Lau}}, \
  and\ \bibinfo {author} {\bibfnamefont {V.}~\bibnamefont {Sokalski}},\
  }\href@noop {} {\bibfield  {journal} {\bibinfo  {journal} {arXiv preprint
  arXiv:1609.04386}\ } (\bibinfo {year} {2016})}\BibitemShut {NoStop}%
\bibitem [{\citenamefont {Di}\ \emph {et~al.}(2015)\citenamefont {Di},
  \citenamefont {Zhang}, \citenamefont {Lim}, \citenamefont {Ng}, \citenamefont
  {Kuok}, \citenamefont {Yu}, \citenamefont {Yoon}, \citenamefont {Qiu},\ and\
  \citenamefont {Yang}}]{di2015direct}%
  \BibitemOpen
  \bibfield  {author} {\bibinfo {author} {\bibfnamefont {K.}~\bibnamefont
  {Di}}, \bibinfo {author} {\bibfnamefont {V.~L.}\ \bibnamefont {Zhang}},
  \bibinfo {author} {\bibfnamefont {H.~S.}\ \bibnamefont {Lim}}, \bibinfo
  {author} {\bibfnamefont {S.~C.}\ \bibnamefont {Ng}}, \bibinfo {author}
  {\bibfnamefont {M.~H.}\ \bibnamefont {Kuok}}, \bibinfo {author}
  {\bibfnamefont {J.}~\bibnamefont {Yu}}, \bibinfo {author} {\bibfnamefont
  {J.}~\bibnamefont {Yoon}}, \bibinfo {author} {\bibfnamefont {X.}~\bibnamefont
  {Qiu}}, \ and\ \bibinfo {author} {\bibfnamefont {H.}~\bibnamefont {Yang}},\
  }\href@noop {} {\bibfield  {journal} {\bibinfo  {journal} {Phys. Rev. Lett.}\
  }\textbf {\bibinfo {volume} {114}},\ \bibinfo {pages} {047201} (\bibinfo
  {year} {2015})}\BibitemShut {NoStop}%
\bibitem [{\citenamefont {Belmeguenai}\ \emph {et~al.}(2015)\citenamefont
  {Belmeguenai}, \citenamefont {Adam}, \citenamefont {Roussign{\'e}},
  \citenamefont {Eimer}, \citenamefont {Devolder}, \citenamefont {Kim},
  \citenamefont {Ch{\'e}rif}, \citenamefont {Stashkevich},\ and\ \citenamefont
  {Thiaville}}]{belmeguenai2015interfacial}%
  \BibitemOpen
  \bibfield  {author} {\bibinfo {author} {\bibfnamefont {M.}~\bibnamefont
  {Belmeguenai}}, \bibinfo {author} {\bibfnamefont {J.-P.}\ \bibnamefont
  {Adam}}, \bibinfo {author} {\bibfnamefont {Y.}~\bibnamefont {Roussign{\'e}}},
  \bibinfo {author} {\bibfnamefont {S.}~\bibnamefont {Eimer}}, \bibinfo
  {author} {\bibfnamefont {T.}~\bibnamefont {Devolder}}, \bibinfo {author}
  {\bibfnamefont {J.-V.}\ \bibnamefont {Kim}}, \bibinfo {author} {\bibfnamefont
  {S.~M.}\ \bibnamefont {Ch{\'e}rif}}, \bibinfo {author} {\bibfnamefont
  {A.}~\bibnamefont {Stashkevich}}, \ and\ \bibinfo {author} {\bibfnamefont
  {A.}~\bibnamefont {Thiaville}},\ }\href@noop {} {\bibfield  {journal}
  {\bibinfo  {journal} {Phys. Rev. B}\ }\textbf {\bibinfo {volume} {91}},\
  \bibinfo {pages} {180405} (\bibinfo {year} {2015})}\BibitemShut {NoStop}%
\bibitem [{\citenamefont {Nembach}\ \emph {et~al.}(2015)\citenamefont
  {Nembach}, \citenamefont {Shaw}, \citenamefont {Weiler}, \citenamefont
  {Ju{\'e}},\ and\ \citenamefont {Silva}}]{nembach2015linear}%
  \BibitemOpen
  \bibfield  {author} {\bibinfo {author} {\bibfnamefont {H.~T.}\ \bibnamefont
  {Nembach}}, \bibinfo {author} {\bibfnamefont {J.~M.}\ \bibnamefont {Shaw}},
  \bibinfo {author} {\bibfnamefont {M.}~\bibnamefont {Weiler}}, \bibinfo
  {author} {\bibfnamefont {E.}~\bibnamefont {Ju{\'e}}}, \ and\ \bibinfo
  {author} {\bibfnamefont {T.~J.}\ \bibnamefont {Silva}},\ }\href@noop {}
  {\bibfield  {journal} {\bibinfo  {journal} {Nat. Phys.}\ }\textbf {\bibinfo
  {volume} {11}},\ \bibinfo {pages} {825–} (\bibinfo {year}
  {2015})}\BibitemShut {NoStop}%
\bibitem [{\citenamefont {Rowan-Robinson}\ \emph {et~al.}(2017)\citenamefont
  {Rowan-Robinson}, \citenamefont {Stashkevich}, \citenamefont {Roussign\'{e}},
  \citenamefont {Belmeguenai}, \citenamefont {Ch\'{e}rif}, \citenamefont
  {Thiaville}, \citenamefont {Hase}, \citenamefont {Hindmarch},\ and\
  \citenamefont {Atkinson}}]{rowan2017interfacial}%
  \BibitemOpen
  \bibfield  {author} {\bibinfo {author} {\bibfnamefont {R.}~\bibnamefont
  {Rowan-Robinson}}, \bibinfo {author} {\bibfnamefont {A.}~\bibnamefont
  {Stashkevich}}, \bibinfo {author} {\bibfnamefont {Y.}~\bibnamefont
  {Roussign\'{e}}}, \bibinfo {author} {\bibfnamefont {M.}~\bibnamefont
  {Belmeguenai}}, \bibinfo {author} {\bibfnamefont {S.}~\bibnamefont
  {Ch\'{e}rif}}, \bibinfo {author} {\bibfnamefont {A.}~\bibnamefont
  {Thiaville}}, \bibinfo {author} {\bibfnamefont {T.}~\bibnamefont {Hase}},
  \bibinfo {author} {\bibfnamefont {A.}~\bibnamefont {Hindmarch}}, \ and\
  \bibinfo {author} {\bibfnamefont {D.}~\bibnamefont {Atkinson}},\ }\href@noop
  {} {\bibfield  {journal} {\bibinfo  {journal} {arXiv preprint
  arXiv:1704.01338}\ } (\bibinfo {year} {2017})}\BibitemShut {NoStop}%
\bibitem [{\citenamefont {Cort{\'e}s-Ortu{\~n}o}\ and\ \citenamefont
  {Landeros}(2013)}]{cortes2013influence}%
  \BibitemOpen
  \bibfield  {author} {\bibinfo {author} {\bibfnamefont {D.}~\bibnamefont
  {Cort{\'e}s-Ortu{\~n}o}}\ and\ \bibinfo {author} {\bibfnamefont
  {P.}~\bibnamefont {Landeros}},\ }\href@noop {} {\bibfield  {journal}
  {\bibinfo  {journal} {J. Phys. Condens. Matter}\ }\textbf {\bibinfo {volume}
  {25}},\ \bibinfo {pages} {156001} (\bibinfo {year} {2013})}\BibitemShut
  {NoStop}%
\bibitem [{\citenamefont {Blake}\ \emph {et~al.}(2007)\citenamefont {Blake},
  \citenamefont {Hill}, \citenamefont {Neto}, \citenamefont {Novoselov},
  \citenamefont {Jiang}, \citenamefont {Yang}, \citenamefont {Booth},\ and\
  \citenamefont {Geim}}]{blake2007making}%
  \BibitemOpen
  \bibfield  {author} {\bibinfo {author} {\bibfnamefont {P.}~\bibnamefont
  {Blake}}, \bibinfo {author} {\bibfnamefont {E.}~\bibnamefont {Hill}},
  \bibinfo {author} {\bibfnamefont {A.~C.}\ \bibnamefont {Neto}}, \bibinfo
  {author} {\bibfnamefont {K.}~\bibnamefont {Novoselov}}, \bibinfo {author}
  {\bibfnamefont {D.}~\bibnamefont {Jiang}}, \bibinfo {author} {\bibfnamefont
  {R.}~\bibnamefont {Yang}}, \bibinfo {author} {\bibfnamefont {T.}~\bibnamefont
  {Booth}}, \ and\ \bibinfo {author} {\bibfnamefont {A.}~\bibnamefont {Geim}},\
  }\href@noop {} {\bibfield  {journal} {\bibinfo  {journal} {Appl. Phys.
  Lett.}\ }\textbf {\bibinfo {volume} {91}},\ \bibinfo {pages} {063124}
  (\bibinfo {year} {2007})}\BibitemShut {NoStop}%
\bibitem [{\citenamefont {Wang}\ \emph {et~al.}(2008)\citenamefont {Wang},
  \citenamefont {Ni}, \citenamefont {Shen}, \citenamefont {Wang},\ and\
  \citenamefont {Wu}}]{wang2008interference}%
  \BibitemOpen
  \bibfield  {author} {\bibinfo {author} {\bibfnamefont {Y.}~\bibnamefont
  {Wang}}, \bibinfo {author} {\bibfnamefont {Z.}~\bibnamefont {Ni}}, \bibinfo
  {author} {\bibfnamefont {Z.}~\bibnamefont {Shen}}, \bibinfo {author}
  {\bibfnamefont {H.}~\bibnamefont {Wang}}, \ and\ \bibinfo {author}
  {\bibfnamefont {Y.}~\bibnamefont {Wu}},\ }\href@noop {} {\bibfield  {journal}
  {\bibinfo  {journal} {Appl. Phys. Lett.}\ }\textbf {\bibinfo {volume} {92}},\
  \bibinfo {pages} {043121} (\bibinfo {year} {2008})}\BibitemShut {NoStop}%
\bibitem [{\citenamefont {Yoon}\ \emph {et~al.}(2009)\citenamefont {Yoon},
  \citenamefont {Moon}, \citenamefont {Son}, \citenamefont {Choi},
  \citenamefont {Park}, \citenamefont {Cha}, \citenamefont {Kim},\ and\
  \citenamefont {Cheong}}]{yoon2009interference}%
  \BibitemOpen
  \bibfield  {author} {\bibinfo {author} {\bibfnamefont {D.}~\bibnamefont
  {Yoon}}, \bibinfo {author} {\bibfnamefont {H.}~\bibnamefont {Moon}}, \bibinfo
  {author} {\bibfnamefont {Y.-W.}\ \bibnamefont {Son}}, \bibinfo {author}
  {\bibfnamefont {J.~S.}\ \bibnamefont {Choi}}, \bibinfo {author}
  {\bibfnamefont {B.~H.}\ \bibnamefont {Park}}, \bibinfo {author}
  {\bibfnamefont {Y.~H.}\ \bibnamefont {Cha}}, \bibinfo {author} {\bibfnamefont
  {Y.~D.}\ \bibnamefont {Kim}}, \ and\ \bibinfo {author} {\bibfnamefont
  {H.}~\bibnamefont {Cheong}},\ }\href@noop {} {\bibfield  {journal} {\bibinfo
  {journal} {Phys. Rev. B}\ }\textbf {\bibinfo {volume} {80}},\ \bibinfo
  {pages} {125422} (\bibinfo {year} {2009})}\BibitemShut {NoStop}%
\bibitem [{\citenamefont {Jackson}(1999)}]{jackson1999classical}%
  \BibitemOpen
  \bibfield  {author} {\bibinfo {author} {\bibfnamefont {J.~D.}\ \bibnamefont
  {Jackson}},\ }\href@noop {} {\emph {\bibinfo {title} {Classical
  electrodynamics}}}\ (\bibinfo  {publisher} {Wiley},\ \bibinfo {year}
  {1999})\BibitemShut {NoStop}%
\bibitem [{\citenamefont {Hubert}\ and\ \citenamefont
  {Sch{\"a}fer}(1998)}]{hubert1998magnetic}%
  \BibitemOpen
  \bibfield  {author} {\bibinfo {author} {\bibfnamefont {A.}~\bibnamefont
  {Hubert}}\ and\ \bibinfo {author} {\bibfnamefont {R.}~\bibnamefont
  {Sch{\"a}fer}},\ }\href@noop {} {\emph {\bibinfo {title} {Magnetic domains:
  the analysis of magnetic microstructure}}}\ (\bibinfo  {publisher} {Springer
  Berlin},\ \bibinfo {year} {1998})\BibitemShut {NoStop}%
\bibitem [{\citenamefont {Loudon}\ and\ \citenamefont
  {Sandercock}(1980)}]{loudon1980analysis}%
  \BibitemOpen
  \bibfield  {author} {\bibinfo {author} {\bibfnamefont {R.}~\bibnamefont
  {Loudon}}\ and\ \bibinfo {author} {\bibfnamefont {J.}~\bibnamefont
  {Sandercock}},\ }\href@noop {} {\bibfield  {journal} {\bibinfo  {journal} {J.
  Phys. C}\ }\textbf {\bibinfo {volume} {13}},\ \bibinfo {pages} {2609}
  (\bibinfo {year} {1980})}\BibitemShut {NoStop}%
\bibitem [{\citenamefont {Stashkevich}\ \emph {et~al.}(2009)\citenamefont
  {Stashkevich}, \citenamefont {Roussign{\'e}}, \citenamefont {Djemia},
  \citenamefont {Ch{\'e}rif}, \citenamefont {Evans}, \citenamefont {Murphy},
  \citenamefont {Hendren}, \citenamefont {Atkinson}, \citenamefont {Pollard},
  \citenamefont {Zayats} \emph {et~al.}}]{stashkevich2009spin}%
  \BibitemOpen
  \bibfield  {author} {\bibinfo {author} {\bibfnamefont {A.~A.}\ \bibnamefont
  {Stashkevich}}, \bibinfo {author} {\bibfnamefont {Y.}~\bibnamefont
  {Roussign{\'e}}}, \bibinfo {author} {\bibfnamefont {P.}~\bibnamefont
  {Djemia}}, \bibinfo {author} {\bibfnamefont {S.~M.}\ \bibnamefont
  {Ch{\'e}rif}}, \bibinfo {author} {\bibfnamefont {P.~R.}\ \bibnamefont
  {Evans}}, \bibinfo {author} {\bibfnamefont {A.~P.}\ \bibnamefont {Murphy}},
  \bibinfo {author} {\bibfnamefont {W.~R.}\ \bibnamefont {Hendren}}, \bibinfo
  {author} {\bibfnamefont {R.}~\bibnamefont {Atkinson}}, \bibinfo {author}
  {\bibfnamefont {R.~J.}\ \bibnamefont {Pollard}}, \bibinfo {author}
  {\bibfnamefont {A.~V.}\ \bibnamefont {Zayats}},  \emph {et~al.},\ }\href@noop
  {} {\bibfield  {journal} {\bibinfo  {journal} {Phys. Rev. B}\ }\textbf
  {\bibinfo {volume} {80}},\ \bibinfo {pages} {144406} (\bibinfo {year}
  {2009})}\BibitemShut {NoStop}%
\bibitem [{\citenamefont {Kranz}\ and\ \citenamefont
  {Hubert}(1963)}]{kranz1963moglichkeiten}%
  \BibitemOpen
  \bibfield  {author} {\bibinfo {author} {\bibfnamefont {J.}~\bibnamefont
  {Kranz}}\ and\ \bibinfo {author} {\bibfnamefont {A.}~\bibnamefont {Hubert}},\
  }\href@noop {} {\bibfield  {journal} {\bibinfo  {journal} {Z. angew. Phys.}\
  }\textbf {\bibinfo {volume} {15}},\ \bibinfo {pages} {220} (\bibinfo {year}
  {1963})}\BibitemShut {NoStop}%
\end{thebibliography}
\end{document}